\documentclass[10pt]{article}
\usepackage[OE]{express}

\begin{document}

\title{Design of an optimized $\text{MoS}_2$-based highly sensitive near-infrared surface plasmon resonance imaging biosensor}

\author{Yi Xu,\authormark{1} Lin Wu,\authormark{2} and Lay Kee Ang\authormark{1,*}}

\address{\authormark{1}SUTD-MIT International Design Center, Singapore University of Technology and Design, Singapore 487372\\
\authormark{2}Institute of High Performance Computing, Agency for Science, Technology, and Research (A*STAR), 1 Fusionopolis Way, \#16-16 Connexis, Singapore 138632}

\email{\authormark{*}ricky\_ang@sutd.edu.sg} 



\begin{abstract}
A surface plasmon resonance imaging biosensor based on $\text{MoS}_2$ deposited on Aluminium substrate is designed for high imaging sensitivity and detection accuracy. The proposed biosensor exhibits better performance than graphene-based biosensor in the near-infrared regime. A high imaging sensitivity of more than 970 $\text{RIU}^{-1}$ is obtained at the wavelength of 1540 nm. The effect of aluminium thickness, number of $\text{MoS}_2$ layers and the refractive index of sensing layer are investigated to obtain an optimized design for high sensor performance. In addition, the sensor performance comparison of $\text{MoS}_2$ and other two-dimensional transition metal dichalcogenide materials based biosensor in the near-infrared regime are also presented. The designed $\text{MoS}_2$ mediated surface plasmon resonance imaging biosensor could provide potential applications in surface plasmon resonance imaging detection of multiple biomolecular interactions simultaneously. 
\end{abstract}
  


\section{Introduction}
Surface plasmon resonance (SPR), in which surface plasmon waves (SPWs) are excited at the metal-dielectric interface, has been widely employed in the sensing applications over the last few decades \cite{homola1999surface,homola2008surface,fan2008sensitive,wijaya2011surface}. Many optical structures have been proposed to excite SPWs \cite{homola1999surface,homola2008surface}, like prism-coupling, waveguide-coupling and grating-coupling. A typical SPW excitation structure is the Kretschmann configuration \cite{kretschmann1968notizen}, where a thin metal film is coated over the base of a prism. The SPWs are excited by a p-polarized light when the propagation constant of the incident light along the metal-dielectric interface matches the propagation constant of SPW \cite{homola1999surface}. The excitation of SPWs depends on the refractive index (RI) of dielectric medium that in contact with the metal thin film. Therefore, SPR can be used to detect the variation of ambient RI. One of the widely used SPR sensing techniques is the angle interrogation scheme, which is a sensitive and robust detection method for SPR sensors. In this approach, a p-polarized monochromatic light is employed to excite SPWs at the metal-dielectric interface, and the reflectance is monitored as a function of the incident angle. However, one limitation of this scheme is its inability to measure diverse sets of samples at a time. SPR imaging sensors \cite{wark2005long,lee2006surface,zeng2017recent,choi2010investigation,wong2014surface}, measure numerous samples in a parallel manner, have been proposed and demonstrated to overcome this limitation. Noted for a SPR imaging sensor, the spatial variations in reflectivity are measured at a fixed incident angle (i.e., no moving parts) due to the ambient RI changes.  

Noble metals, like gold and silver, are usually employed to excite SPWs in SPR sensors. However, gold-based SPR sensor shows a broader SPR curve which degrades the detection accuracy. In contrast, aluminium (Al) is a promising material for plasmonics application \cite{knight2013aluminum,king2015fano,li2016aluminum,maharana2013ultrasensitive,maharana2014performance} not only because its narrow SPR curve, tunable plasmon resonant from visible to ultraviolet regime, but also for low cost. However, Al is prone to oxidation which will decrease the sensor performances. Recently, two-dimensional (2D) nanomaterials, like graphene, have been proposed to inhibit the oxidation of Al thin film in SPR sensors \cite{maharana2013ultrasensitive,maharana2014performance}. Although coating Al thin film with graphene layers will decrease the sensor sensitivity, the protected SPR sensor can still obtain an exceptionally high sensitivity due to the atomic thickness of the 2D nonmaterial. 

In addition to the graphene-based SPR sensor \cite{maharana2013ultrasensitive,maharana2014performance,choi2011graphene,wu2010highly,ferrari2015science}, 2D transition metal dichalcogenides (TMDCs) such as Molybdenum disulfide ($\text{MoS}_2$) have also been widely employed in sensing applications \cite{perkins2013chemical,zeng2015graphene,mishra2016graphene,rao2015comparative,ouyang2017two,ouyang2016sensitivity,kalantar2015biosensors,sarkar2014mos2,lee2014two}. Compared with graphene, monolayer $\text{MoS}_2$ has a higher optical absorption efficiency ($\sim$5\%) \cite{lopez2013ultrasensitive}, which provides promising applications in various optoelectronic nanodevices, such as photodetectors with a high responsivity of $5 \times 10^8 \text{AW}^{-1}$ \cite{roy2013graphene}. The nonzero tunable band gap of $\text{MoS}_2$ \cite{mak2010atomically} makes it an attractive candidate for future nanoelectronic devices as well as biosensors \cite{lembke2015single,wang2012electronics}. For example, the nonzero bandgap of $\text{MoS}_2$ can be utilized to fabricate an ultrasensitive field-effect transistor biosensor based on $\text{MoS}_2$, while the zero bandgap in graphene limits the sensitivity of graphene-based field-effect transistor biosensor \cite{sarkar2014mos2}. In addition, the hydrophobic nature of $\text{MoS}_2$ allows it to be used in biosensors as a recognition layer which exhibits high affinity to biomolecules absorption \cite{lee2014two,farimani2014dna}. All these exciting properties make $\text{MoS}_2$ a highly potential candidate for biosensing applications. Taking the advantages of SPR imaging sensor, Al and $\text{MoS}_2$, a highly sensitive and accurate SPR imaging biosensor based on $\text{MoS}_2$ deposited on Al thin film has been designed in this paper. The $\text{MoS}_2$ layer in our designed configuration serves two purposes, as the protective layer of Al thin film and as the recognition layer to capture the biomolecules. We first compared the sensor performance of our design to the graphene-based SPR imaging sensor in the visible and near-infrared regime, which shows that $\text{MoS}_2$-based sensor has a better performance in the near-infrared region. By focusing in the near-infrared regime, we have studied various designed parameters in details in order to obtain an optimized performance including the effects of multiple layers of $\text{MoS}_2$, thickness of the Al film, and RI of the sensing layer. In addition, the performances of other 2D TMDCs based biosensors have also compared at the end. 

\section{Theoretical Model}
\begin{figure}[thpb]
      \centering
      \includegraphics[scale=0.35]{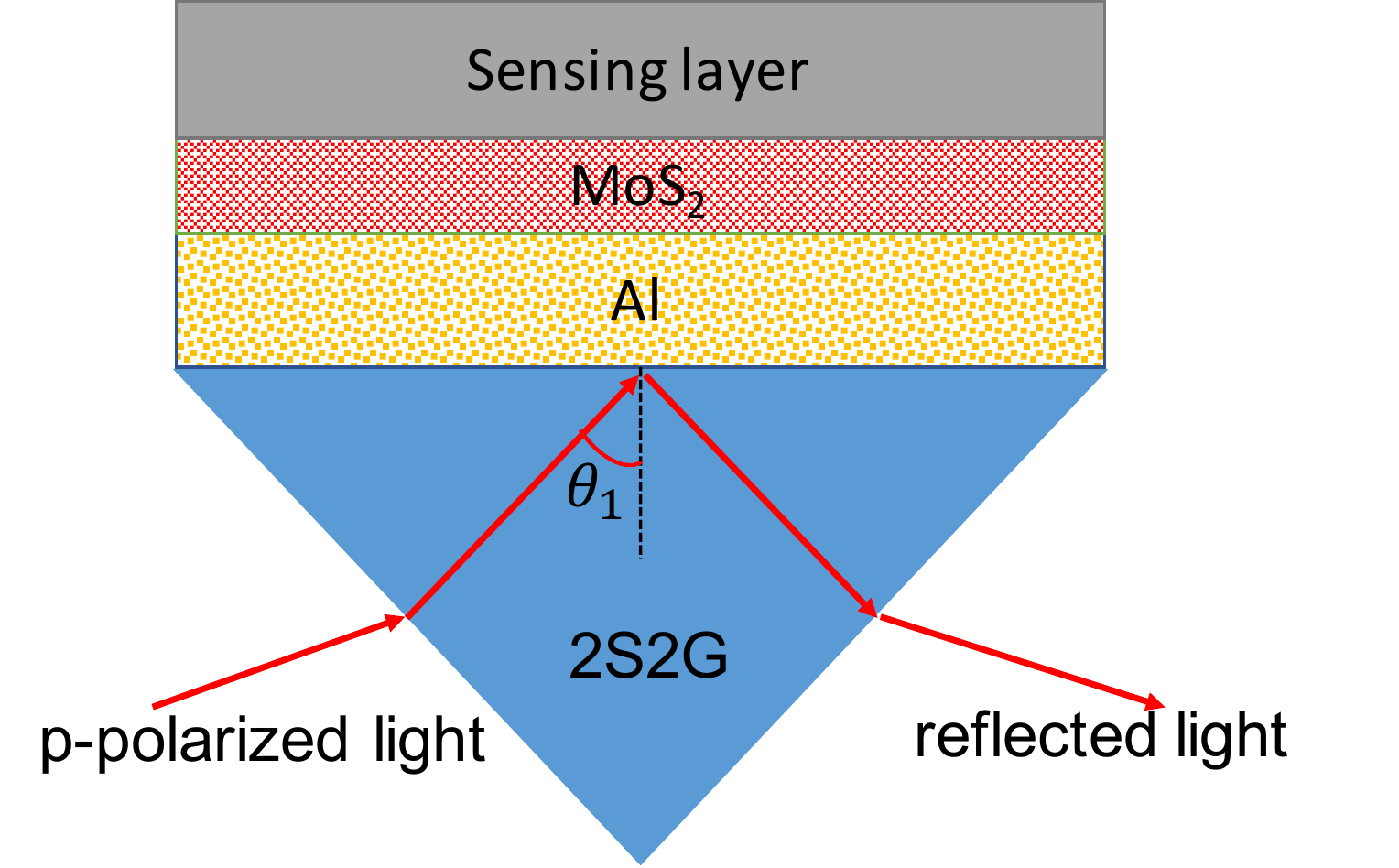}     
      \caption{Schematic of $\text{MoS}_2$-based SPR imaging sensor.}
\label{fig1}
\end{figure}

The Kretschmann configuration is employed to design our proposed biosensor structure, as shown in Fig. \ref{fig1}. In the proposed design, $\text{MoS}_2$ coated Al thin film is attached to a chalcogenide (2S2G) glass prism, which is a promising candidate for the design of SPR sensor due to its broad operating window (from visible to near-infrared regime) and high RI. A p-polarized light with a fixed wavelength is incident at one side of the prism with a fixed incident angle, while the reflected light is collected on the other side. 

The wavelength-dependent RI of the 2S2G prism is given by \cite{maharana2012chalcogenide}: 
\begin{equation}
n_{\text{2S2G}}=2.24047+\frac{2.693\times10^{-2}}{\lambda^2} + \frac{9.08\times10^{-3}}{\lambda^4},
\end{equation}
where the wavelength $\lambda$ is given in $\mu m$. The RI of Al is given by 
\begin{equation}
n_{\text{Al}}=\left(1-\frac{\lambda^2\lambda_c}{\lambda^{2}_p(\lambda_c+i\lambda)}\right)^{1/2},
\end{equation}
according to the Drude-Lorentz model \cite{maharana2014performance}. Here,    $\lambda_p$ (=$1.0657\times10^{-7}$ m) and $\lambda_c$ (=$2.4511\times10^{-5}$ m) is the plasma wavelength and collision wavelength of Al, respectively. The thickness of monolayer $\text{MoS}_2$ is 0.65 nm, and its RI in visible and near-infrared region is shown in Table \ref{tab:table1} \cite{ouyang2017two}. The RI of the sensing layer is initially set to $n_s=1.330$. 
\begin{table}[thpb]
\centering
\caption{\label{tab:table1} RI of Monolayer $\text{MoS}_2$ in the Near-Infrared Region \cite{ouyang2017two} }
\begin{tabular}{cc}
\hline\hline
\textrm{wavelength}&
\textrm{RI}\\
 $\lambda$=633 nm & $5.0805+i1.1732$\\
 $\lambda$=785 nm & $4.6348+i0.1163$ \\
 $\lambda$=904 nm & $4.7261+i0.1346$ \\
 $\lambda$=1150 nm & $4.4317+i0.0721$\\
 $\lambda$=1540 nm & $4.2374 + i0.0325$ \\
 \hline\hline
\end{tabular}
\end{table}

To obtain the reflectance of the sensor configuration, a generalized N-layer model \cite{yamamoto2002surface} was employed, and the reflectance $R$ for the p-polarized incident light is given by
\begin{equation}
R=\left| \frac{(M_{11}+M_{12}q_N)q_1 - (M_{21}+M_{22}q_N)}{(M_{11}+M_{12}q_N)q_1 + (M_{21}+M_{22}q_N)}\right|^2,
\label{eq:R}
\end{equation}
with
\begin{equation}
M=\left[\begin{array}{cc}
M_{11} & M_{12} \\
M_{21} & M_{22}
\end{array}\right]=\prod_{k=2}^{N-1}M_k,
\end{equation}
and
\begin{equation}
M_k=\left[\begin{array}{cc}
\cos \beta_k & -i (\sin \beta_k)/q_k \\
-i q_k \sin \beta_k & \cos \beta_k
\end{array}\right].
\end{equation}

We denote 
\begin{equation}
\beta_k=\frac{2\pi d_k}{\lambda}\left(n_k^2-n_1^2\sin^2\theta_1\right),
\end{equation}
and 
\begin{equation}
q_k=\frac{\left(n_k^2-n_1^2\sin^2\theta_1\right)^{1/2}}{n_k^2},
\end{equation}
where $n_k$ and $d_k$ are respectively the RI and thickness of the $k$th layer with $k=2$ to $N-1$. The first layer ($k=1$) is the 2S2G prism, and the last layer ($k=N$) is the sensing layer. $\theta_1$ is the incident angle at the prism-Al interface, and $\lambda$ is the wavelength of the p-polarized incident light. 

A variation of the sensing layer RI ($n_s$) will cause a change in the reflectance $R$, and the imaging sensitivity of the SPR sensor is defined as 
\begin{equation}
S=\frac{dR}{dn_s},
\end{equation}
where the reflectance $R$ is given in Eq. (\ref{eq:R}). Besides the imaging sensitivity, another important parameter for the sensor performance is the full width at half maximum (FWHM) of the reflectance curve, which describes the detection accuracy of the sensor. To achieve a excellent performance imaging sensor, we note that the sensor should exhibits high imaging sensitivity and low FWHM (i.e., high detection accuracy).  

\section{Results and discussion}
\begin{figure}[thpb]
      \centering
      \includegraphics[scale=0.21]{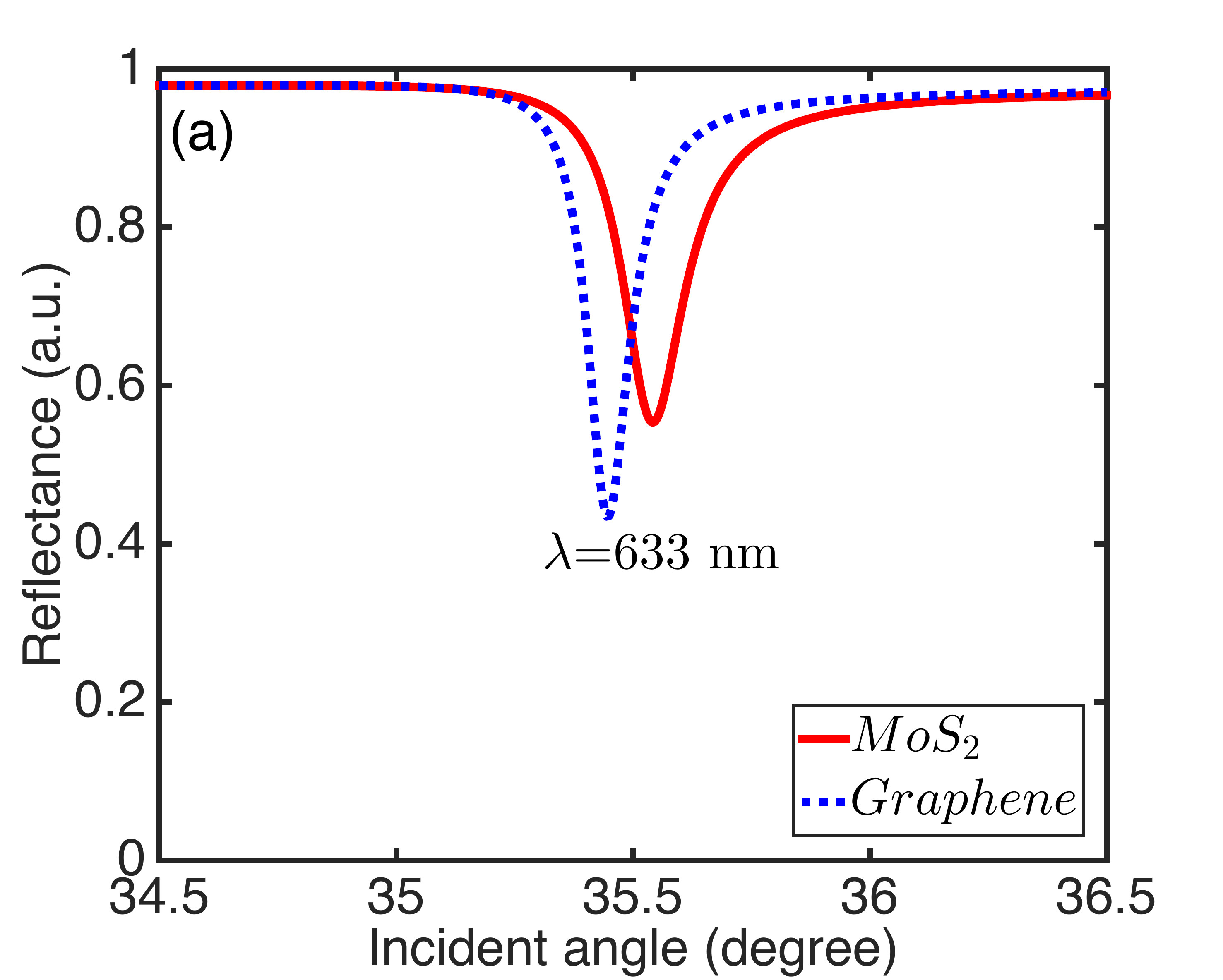}
      \includegraphics[scale=0.21]{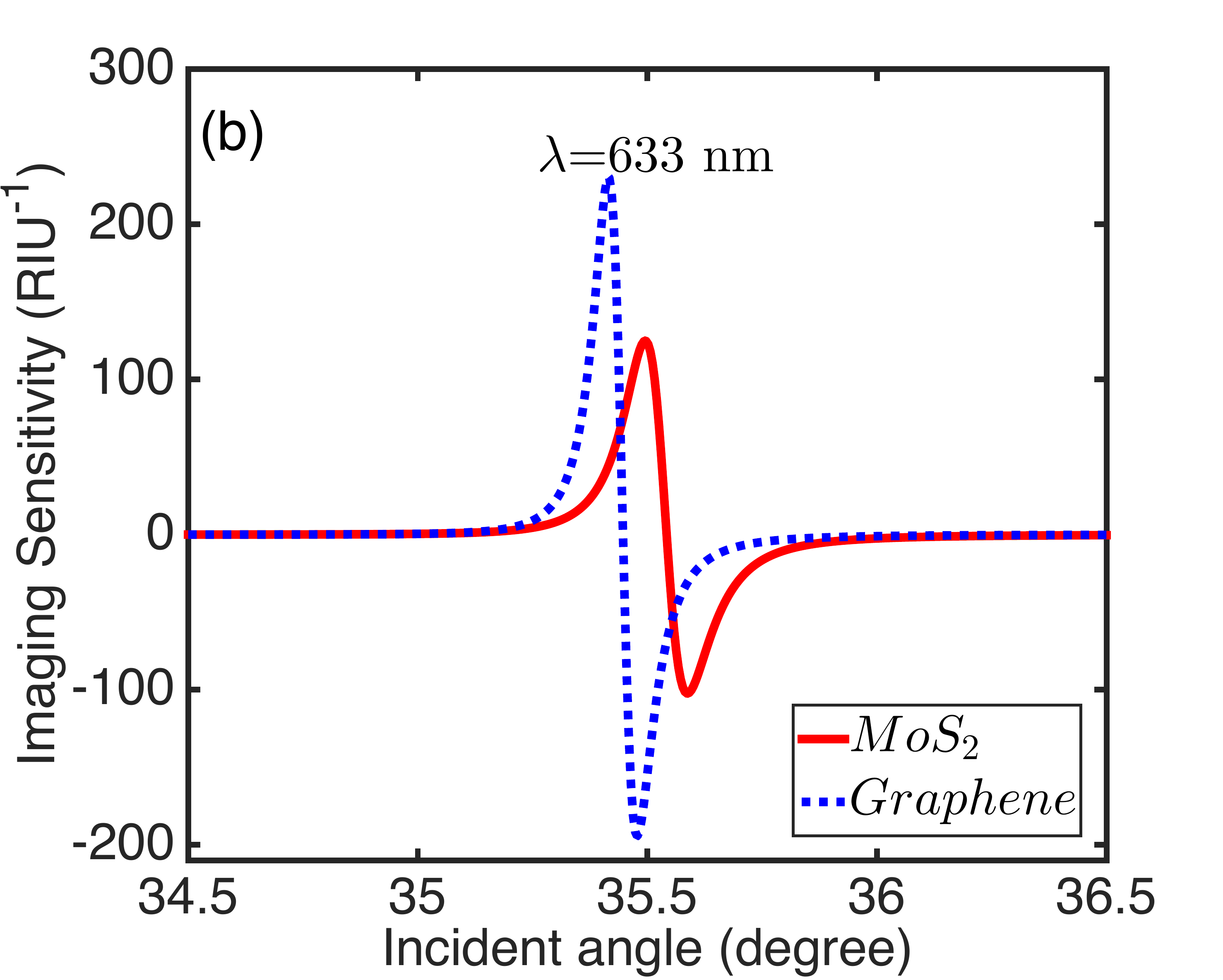}\\ 
      \includegraphics[scale=0.21]{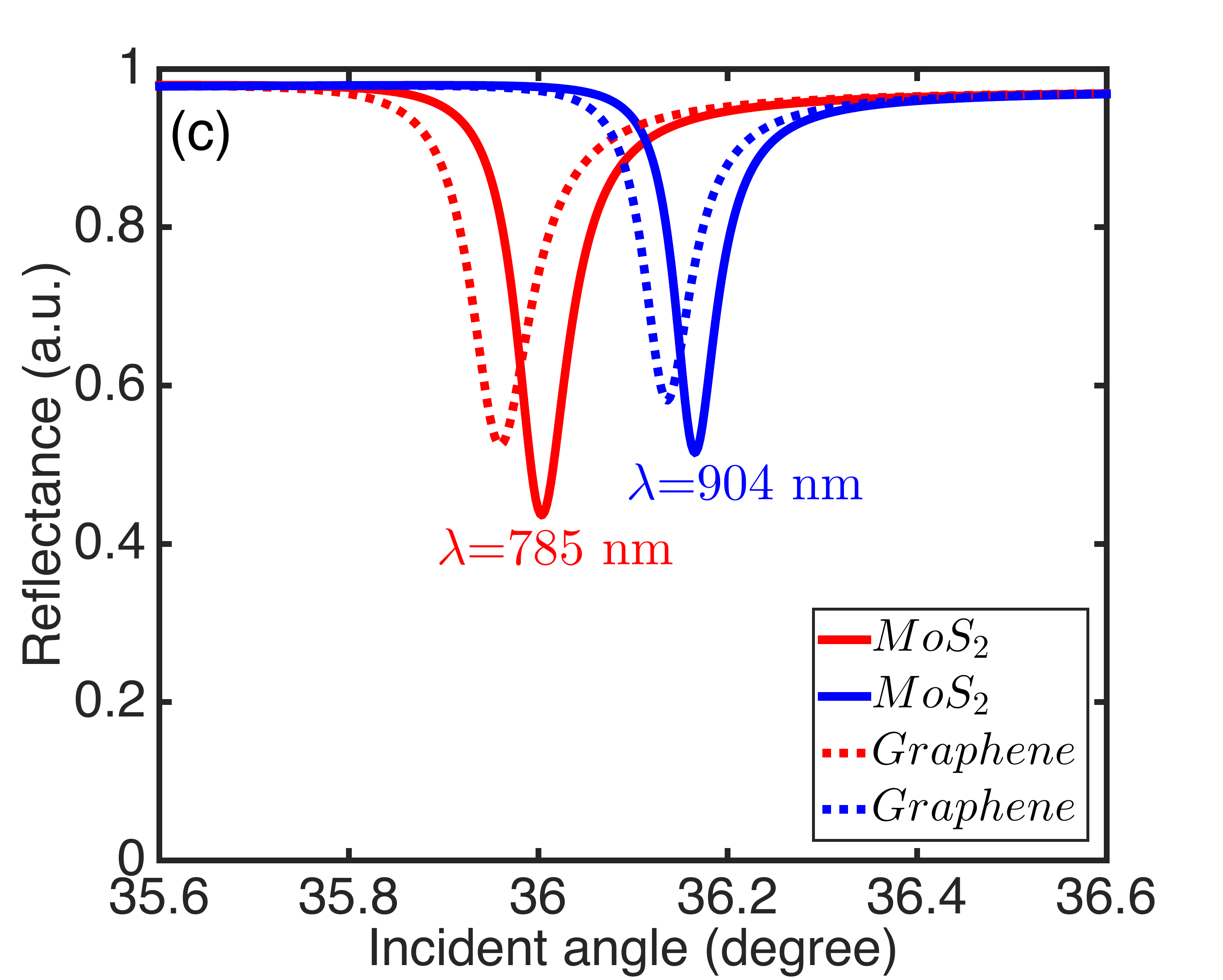}
      \includegraphics[scale=0.21]{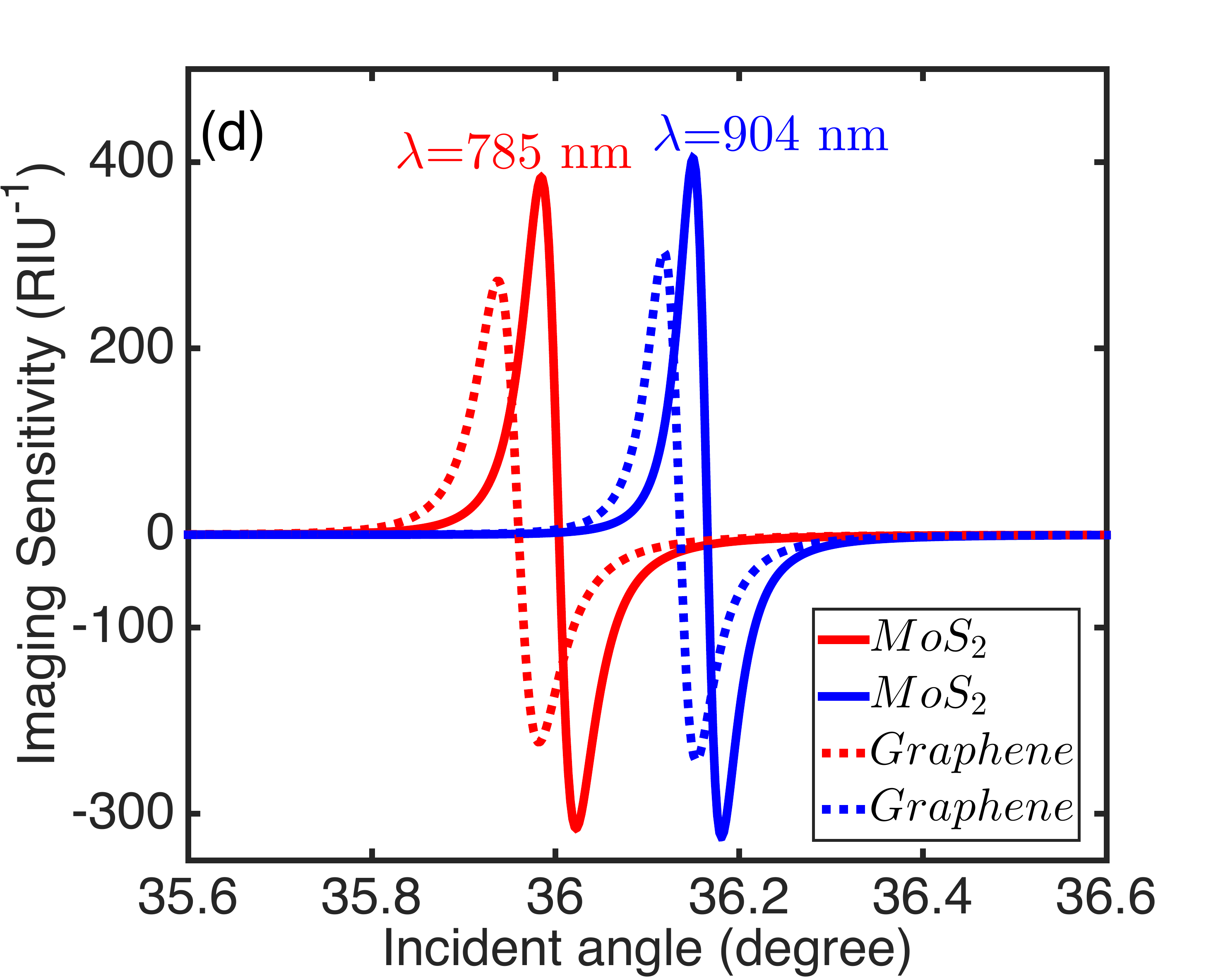}         
      \caption{Reflectance (a)(c) and imaging sensitivity (b)(d) for monolayer $\text{MoS}_2$ and graphene-based sensor at $\lambda=633$ nm, 785 nm and 904 nm. The Al thickness is 50 nm.}
\label{fig2}
\end{figure}

The reflectance and imaging sensitivity for two sensors: a monolayer $\text{MoS}_2$-based sensor and graphene-based sensor, both deposited on a 50 nm thick Al film are shown in  Fig. \ref{fig2} for both visible ($\lambda=633$ nm) and near-infrared ($\lambda=785$ nm and 904 nm) wavelengths. The RI of graphene in the visible and near-infrared region is taken from Ref. \cite{weber2010optical}. In the visible region ($\lambda=633$ nm), graphene-based sensor shows a narrow SPR curve in comparison with that of $\text{MoS}_2$-based sensor ($0.1082^{\text{o}}$ vs. $0.1591^{\text{o}}$), as shown in Fig. \ref{fig2}(a). For the imaging sensitivity, it is found that the sensitivity exhibits a positive peak as well as a negative peak (see Fig. \ref{fig2}(b) and (d)). We only considered the positive peak imaging sensitivity since its magnitude is higher than that of negative peak (see Fig. \ref{fig2}(b) and (d)). In the following, the imaging sensitivity means the positive peak imaging sensitivity. The imaging sensitivity of graphene-based sensor (230.4 $\text{RIU}^{-1}$) is higher than the sensitivity for $\text{MoS}_2$-based imaging sensor (124.8 $\text{RIU}^{-1}$) at $\lambda=633$ nm. However, at the near-infrared region ($\lambda=785$ nm and 904 nm), the imaging sensitivity of $\text{MoS}_2$-based sensor is higher than that of graphene-based sensor (see Fig. \ref{fig2}(d) and Table \ref{tab:table2}). The FWHM of $\text{MoS}_2$-based sensor in the near-infrared region is smaller than that of graphene-based sensor (see Table \ref{tab:table2}). Thus the findings of high imaging sensitivity and low FWHM have indicated that $\text{MoS}_2$ based sensor has better performance than graphene-based sensor in the near-infrared region. The trends also indicates its better performance may be extended to even longer wavelengths, and thus we will focus on three longer wavelengths (785 nm, 1150 nm and 1540 nm) in the optical near-infrared regime in the subsequent studies.

\begin{table*}[thpb]
\centering
\caption{\label{tab:table2}%
Comparison of sensor performances for monolayer $\text{MoS}_2$- and graphene-based sensor in the visible and near-Infrared Region.}
\begin{tabular}{ccccc}
\hline\hline
 &\multicolumn{2}{c}{Imaging Sensitivity ($\text{RIU}^{-1}$)}&\multicolumn{2}{c}{FWHM (degree)}\\
 wavelength & $\text{MoS}_2$-based  & graphene-based & $\text{MoS}_2$-based & graphene-based\\ \hline
 $\lambda=633$ nm & 124.8 & 230.4 & 0.1591 & 0.1082 \\
 $\lambda=785$ nm & 383.4 & 272.7 & 0.0657 & 0.0772 \\
 $\lambda=904$ nm & 403.9 & 303.6 & 0.0538 & 0.0613 \\
\hline\hline
\end{tabular}
\end{table*}

\begin{figure}[thpb]
      \centering
      \includegraphics[scale=0.21]{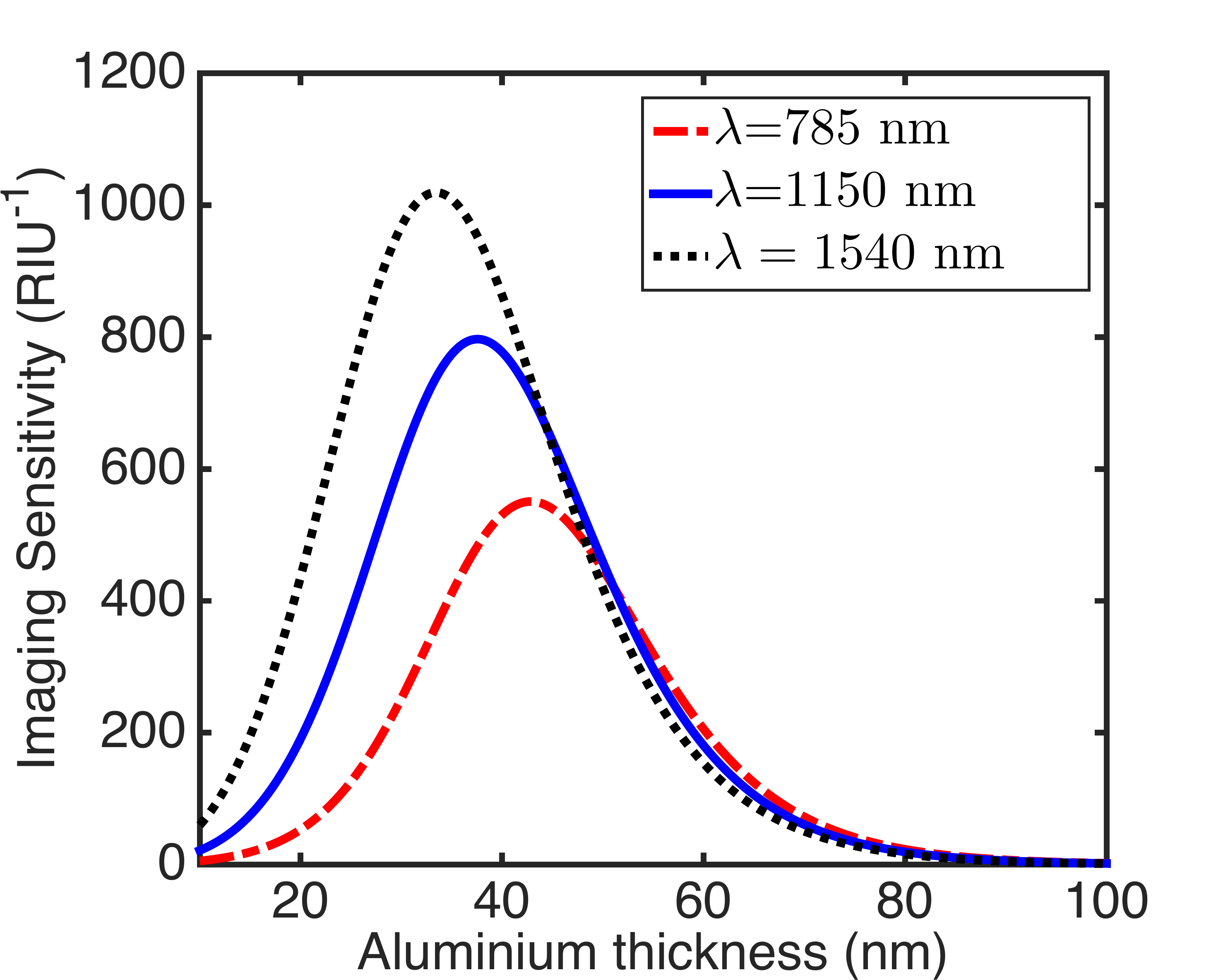}      
      \caption{Imaging sensitivity as a function of the Al thickness without $\text{MoS}_2$ coated.}
\label{fig3}
\end{figure}

Before investigating the sensor performance of $\text{MoS}_2$-based biosensor at longer wavelength, we first optimize the Al thickness since the thickness of plasmonics supporting material is an important parameter for the SPR sensor performance. The imaging sensitivity as a function of the Al thickness (without coated $\text{MoS}_2$) at wavelengths $\lambda=785$ nm, 1150 nm and 1540 nm is shown in Fig. \ref{fig3}. It can be seen from Fig. \ref{fig3} that the optimized Al thickness decreases with the increasing wavelength. Here, we consider the optimized Al thickness is 38 nm, at which at least $90\%$ of the maximum imaging sensitivity's value can be obtained for all three wavelengths. 
 
\begin{figure}[thpb]
      \centering
      \includegraphics[scale=0.21]{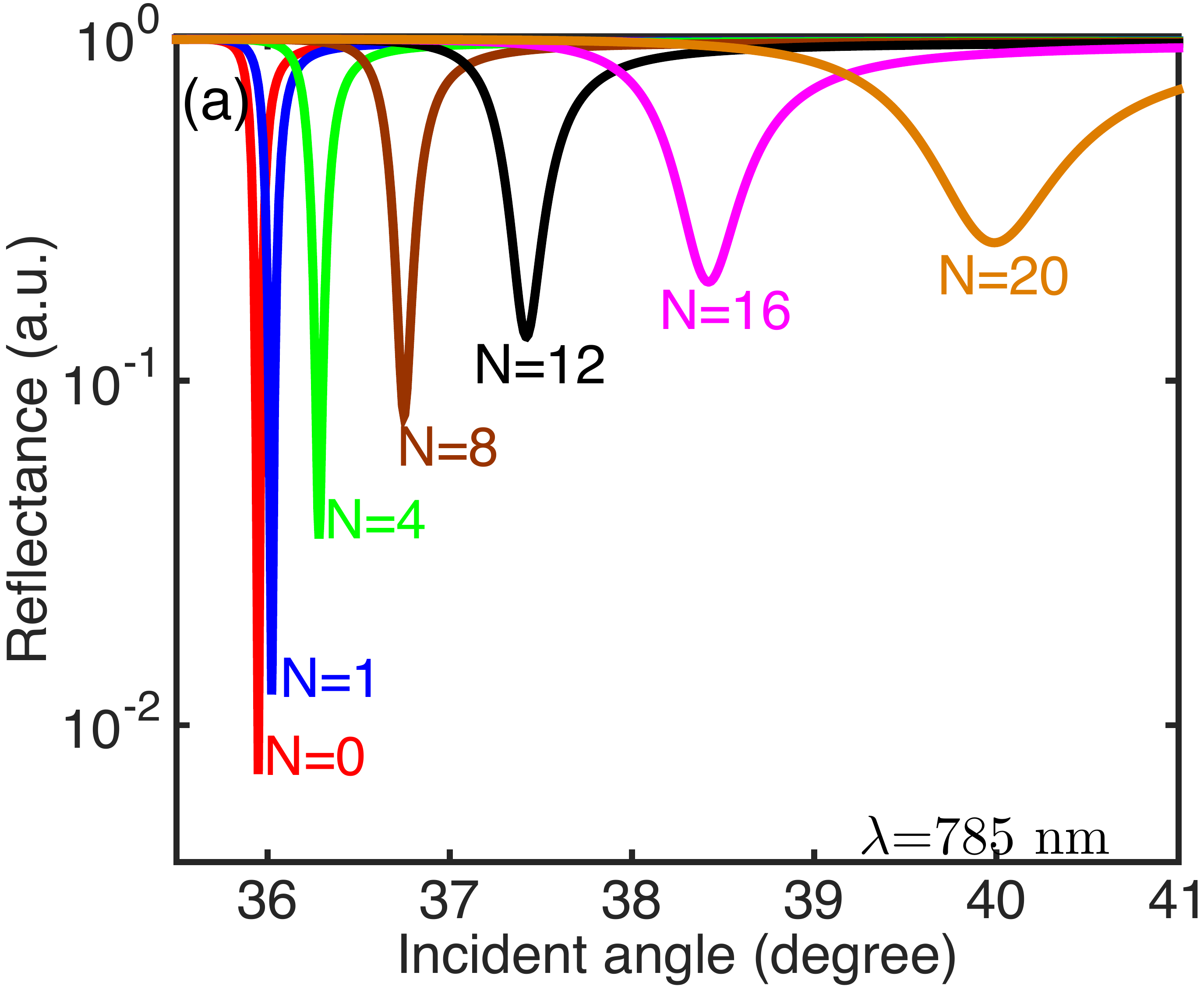}
      \includegraphics[scale=0.21]{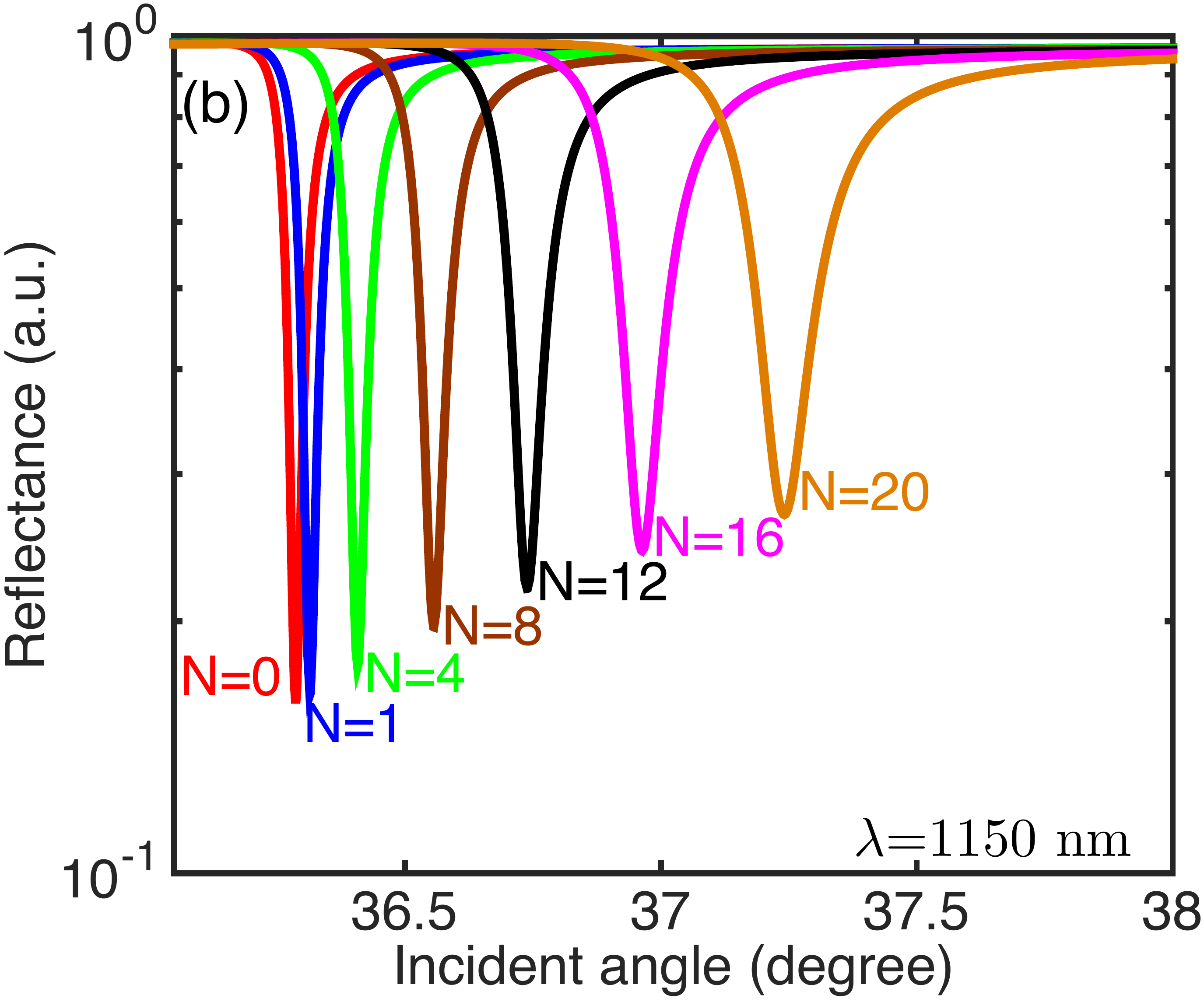}\\
      \includegraphics[scale=0.21]{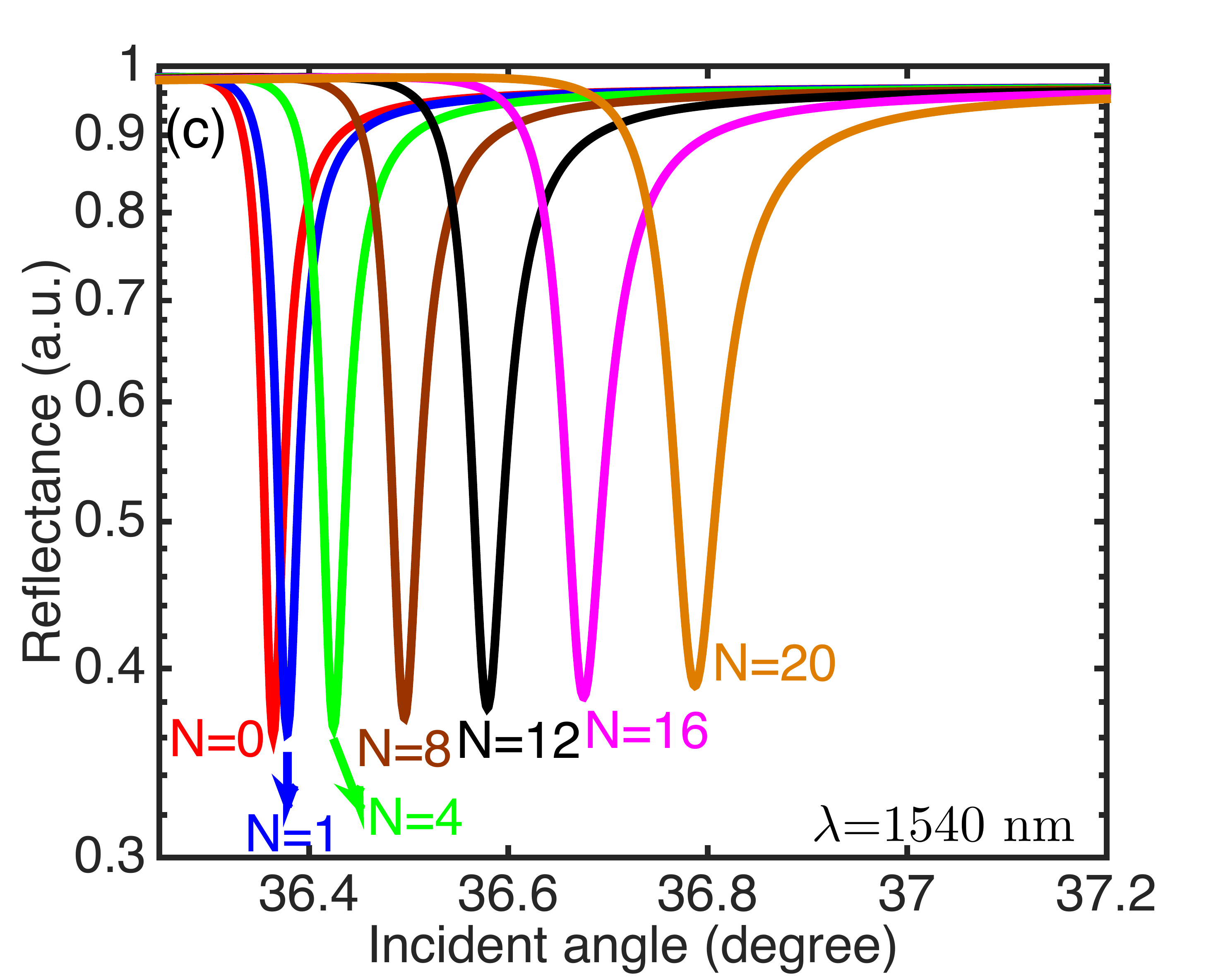}
      \includegraphics[scale=0.21]{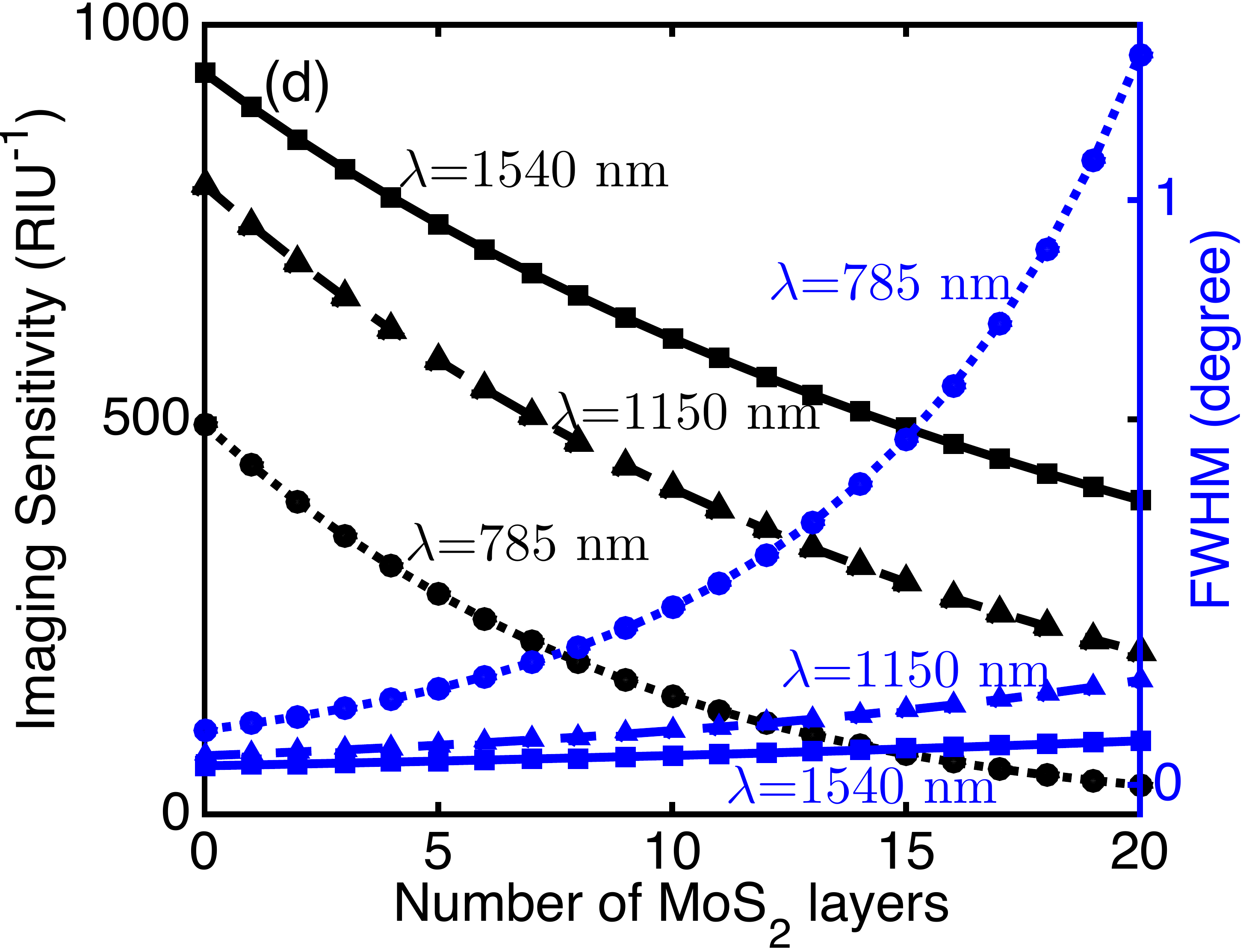}               
      \caption{Reflectance as a function of incident angle for multiple $\text{MoS}_2$-based SPR imaging sensor at wavelength (a) $\lambda=785$ nm, (b) $\lambda=1150$ nm and (c) $\lambda=1540$ nm. (d) Imaging sensitivity and FWHM for SPR imaging sensor with multiple $\text{MoS}_2$ layers at different wavelengths. The Al thickness is 38 nm.}
\label{fig4}
\end{figure}

The effect of multiple $\text{MoS}_2$ layers on the reflectance and sensor performance at three different wavelengths is shown in Fig. \ref{fig4}. It is found that the resonance curves become shallower and broader (larger FWHM, see Fig. \ref{fig4}(d)) with the number of $\text{MoS}_2$ layers increases, which is a result of the increased surface plasmon damping. For example, with monolayer $\text{MoS}_2$ coated sensor, the FWHM is $0.105^{\text{o}}$ at $\lambda=785$ nm, while FWHM is $0.0532^{\text{o}}$ at $\lambda=1150$ nm and $0.0343^{\text{o}}$ at $\lambda=1540$ nm. When the Al thin film is coated with 20 layers $\text{MoS}_2$, the FWHM becomes quite large at $\lambda=785$ nm ($1.249^{\text{o}}$), while the FWHM at longer wavelength is still small ($0.1791^{\text{o}}$ at $\lambda=1150$ nm and $0.0755^{\text{o}}$ at $\lambda=1540$ nm). In addition, the resonance angle increases with the number of $\text{MoS}_2$ layers. For the imaging sensitivity, it decreases with the increasing number of $\text{MoS}_2$ layers (see Fig. \ref{fig4}(d)) due to the increased loss within the $\text{MoS}_2$ layers. As mentioned before, Al is susceptible to oxidation which deceases the sensor performances, whereas $\text{MoS}_2$ layers deposited on Al thin film can be utilized to inhibit the oxidation of Al. Although coating with $\text{MoS}_2$ decreases the imaging sensitivity, the designed biosensors with monolayer or bilayer $\text{MoS}_2$ can still provide exceptional sensitivities. For monolayer $\text{MoS}_2$-based sensor, the imaging sensitivity is 442 $\text{RIU}^{-1}$ at $\lambda=785$ nm, 745.8 $\text{RIU}^{-1}$ at $\lambda=1150$ nm and 895.6 $\text{RIU}^{-1}$ at $\lambda=1540$ nm. It is noted that the proposed biosensor provides a relative high sensitivity of 397.8 $\text{RIU}^{-1}$ at $\lambda=1540$ nm with 20 layers $\text{MoS}_2$, while it is only 36.96 $\text{RIU}^{-1}$ at $\lambda=785$ nm. Higher sensitivity and smaller FWHM are obtained at higher wavelength, which is consistent with the results shown in Fig. \ref{fig2}. 

\begin{figure}[thpb]
      \centering
      \includegraphics[scale=0.185]{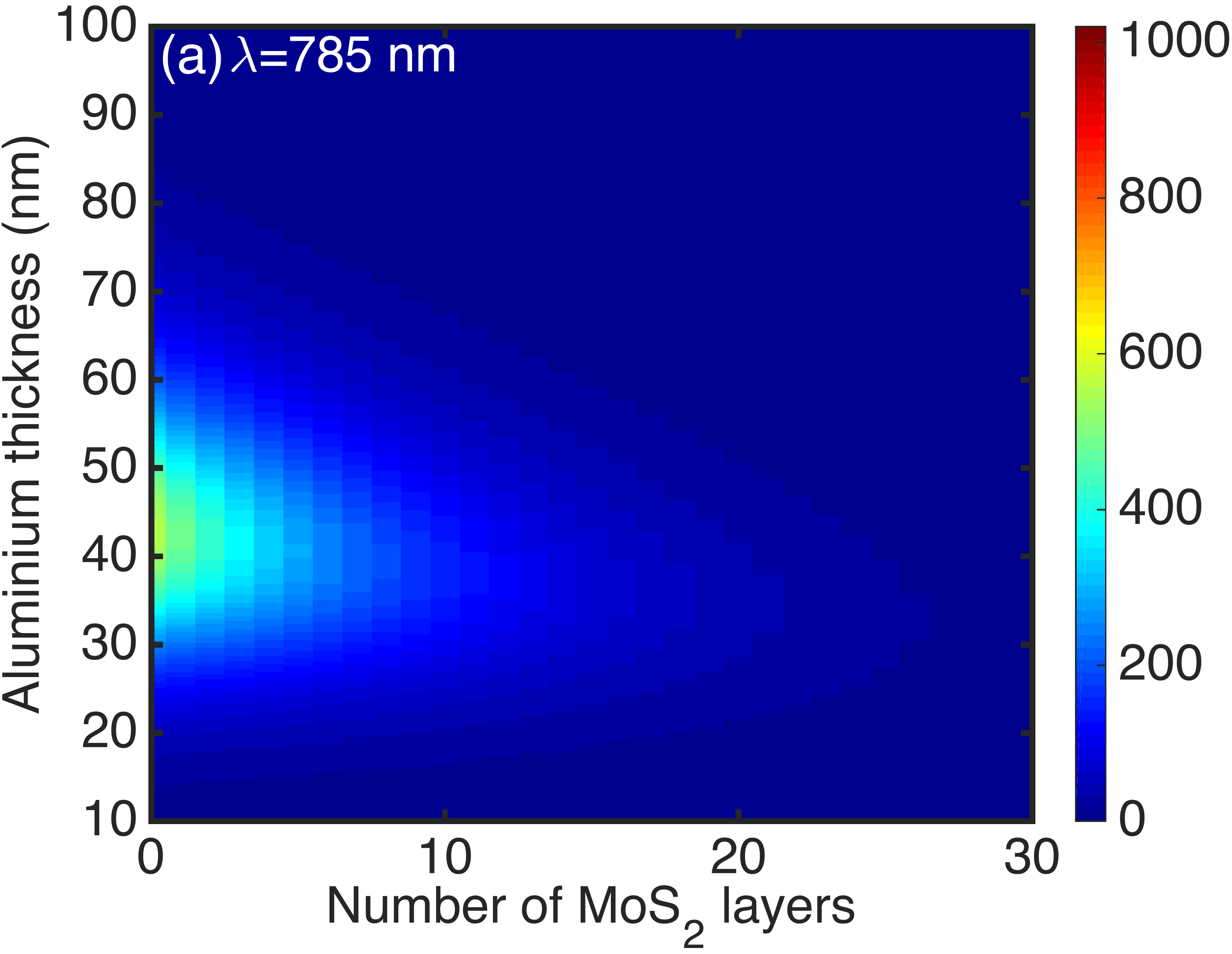} \\  
      \includegraphics[scale=0.18]{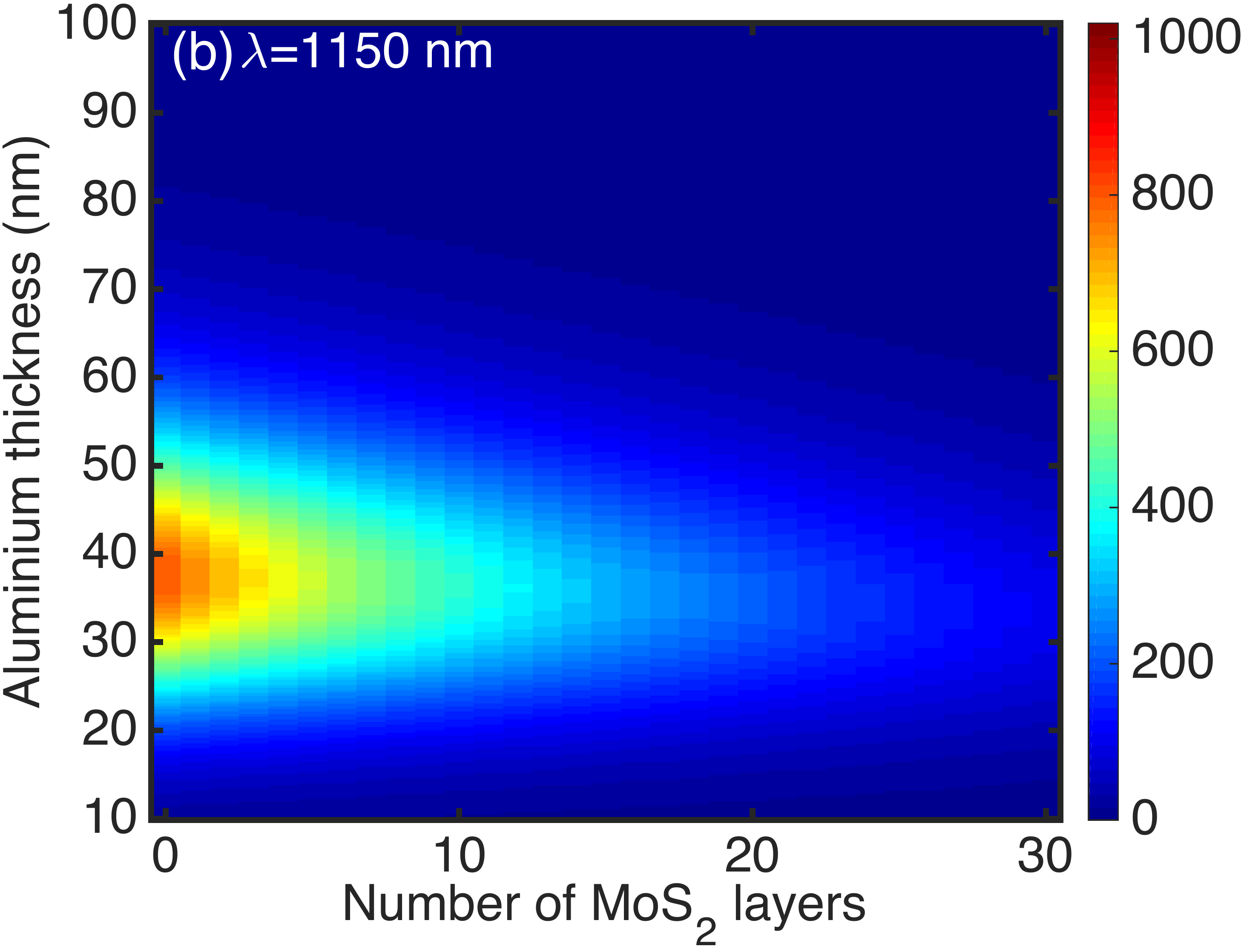}\\
      \includegraphics[scale=0.18]{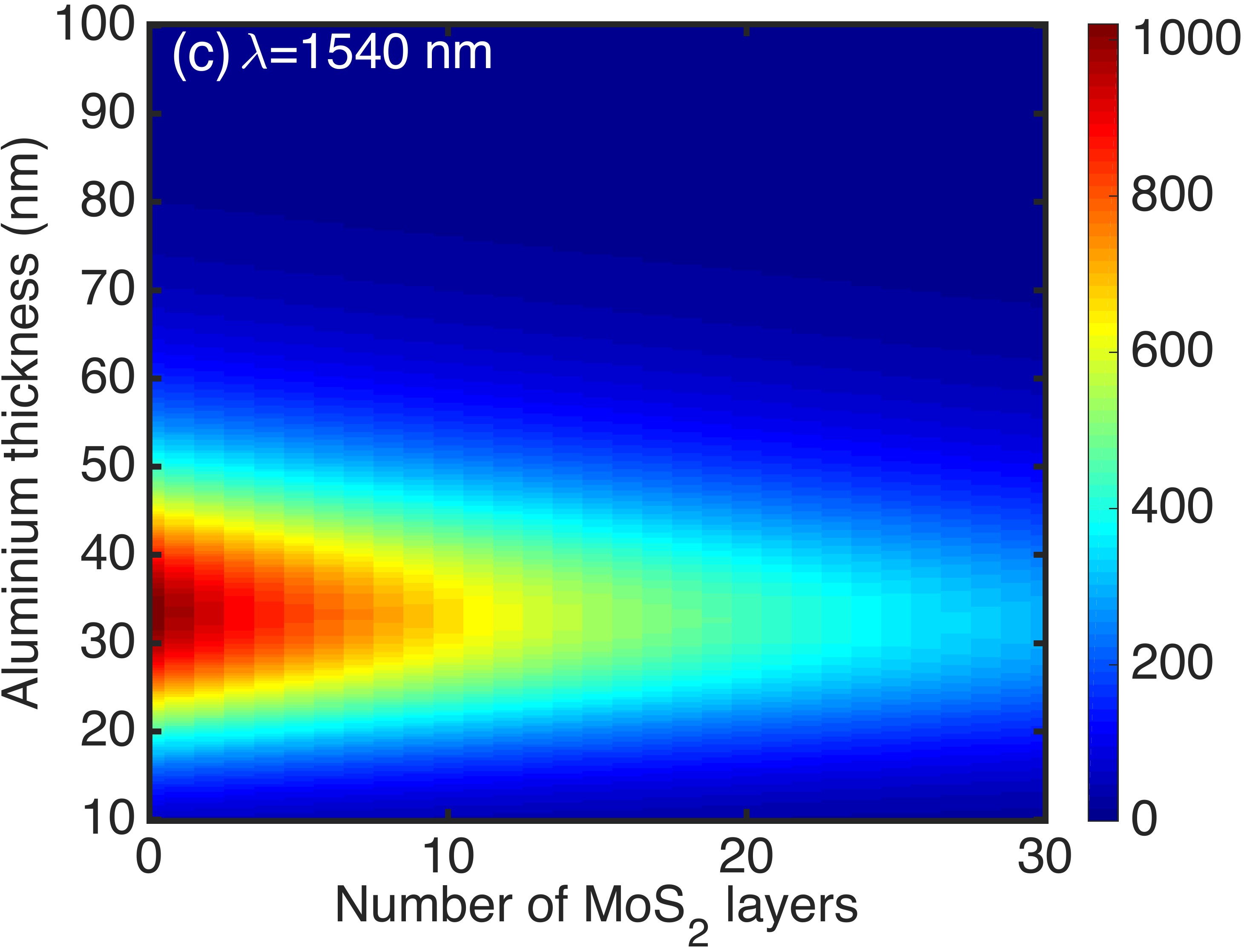}    
      \caption{Imaging sensitivity as a function of the number of $\text{MoS}_2$ layers and Al thickness at (a) $\lambda=785$ nm, (b) $\lambda=1150$ nm and (c) $\lambda=1540$ nm.}
\label{fig5}
\end{figure}

To optimize the design of $\text{MoS}_2$-on-Al SPR imaging sensor, Fig. \ref{fig5} shows the imaging sensitivity as a function of the number of $\text{MoS}_2$ layers and Al thickness at three different wavelengths: 785 nm,  1150 nm and 1540 nm. For monolayer $\text{MoS}_2$-based sensor, highest imaging sensitivity of $\sim$484 $\text{RIU}^{-1}$ is obtained with Al thickness around 42.5 nm at $\lambda=785$ nm, $\sim$747 $\text{RIU}^{-1}$ with Al thickness $\sim37.5$ nm for $\lambda=1150$ nm, and $\sim 974$ $\text{RIU}^{-1}$ with Al thickness $\sim33.5$ nm at wavelength $\lambda=1540$ nm. This indicates that the proposed sensor not only protects the Al from oxidation, but also provides ultrahigh imaging sensitivities. It can be seen from Fig. \ref{fig5} that the designed sensor shows significantly high imaging sensitivity for few $\text{MoS}_2$ layers with 35-50 nm Al thickness at wavelength $\lambda=785$ nm, 30-45 nm Al thickness at $\lambda=1150$ nm, and 25-45 nm Al thickness at $\lambda=1540$ nm. It is noted that the proposed sensor with 15 layers $\text{MoS}_2$ at $\lambda=1150$ nm can still exhibits sensitivity of more than 300 $\text{RIU}^{-1}$ with Al thickness around 35 nm. Even for Al film coated with 30 $\text{MoS}_2$ layers, the proposed sensor can provide an imaging sensitivity of $\sim$300 $\text{RIU}^{-1}$ (more than 290 $\text{RIU}^{-1}$) at wavelength $\lambda=1540$ nm with Al thickness around 30 nm. 

\begin{figure}[thpb]
      \centering
      \includegraphics[scale=0.18]{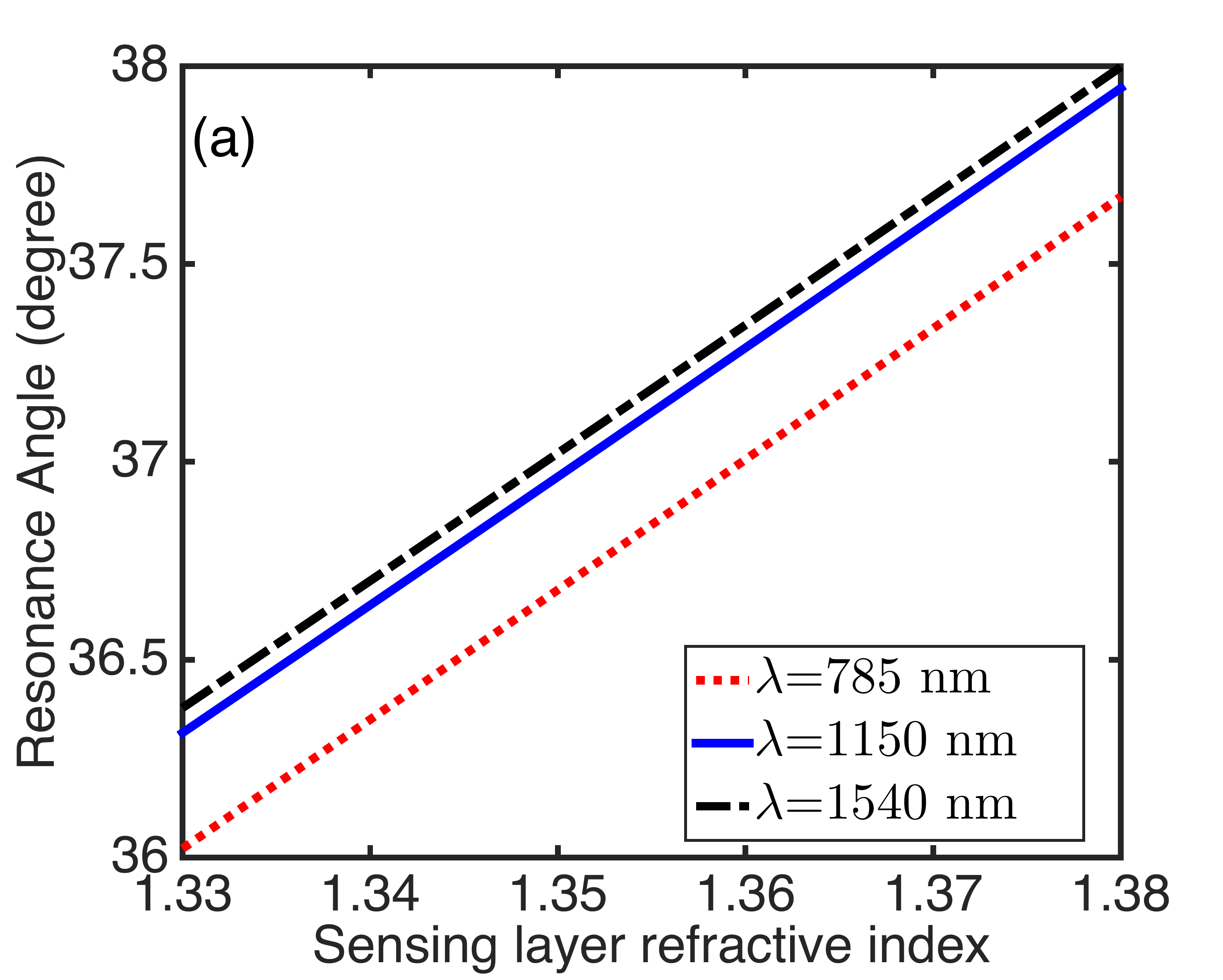}\\
      \includegraphics[scale=0.18]{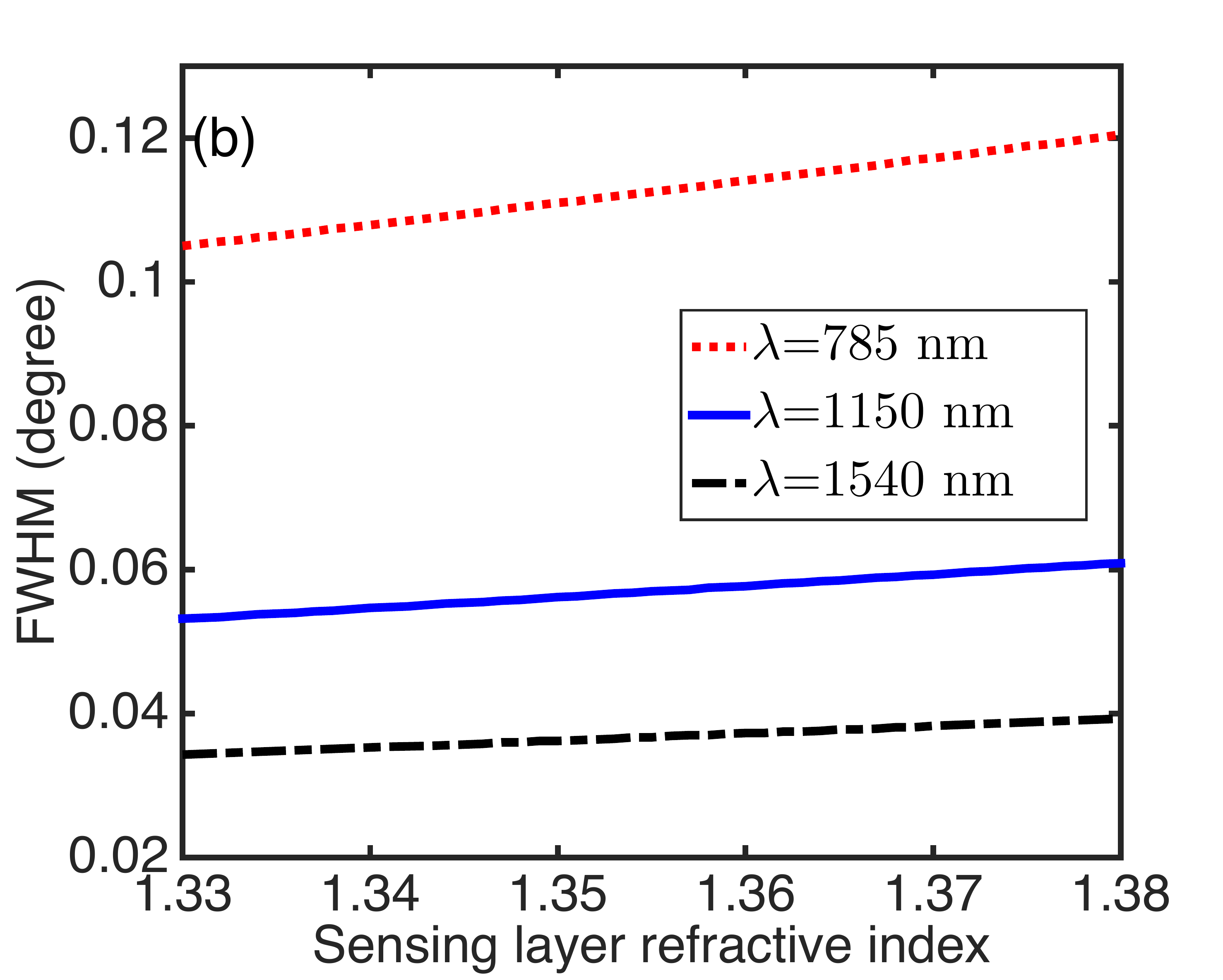}\\
      \includegraphics[scale=0.18]{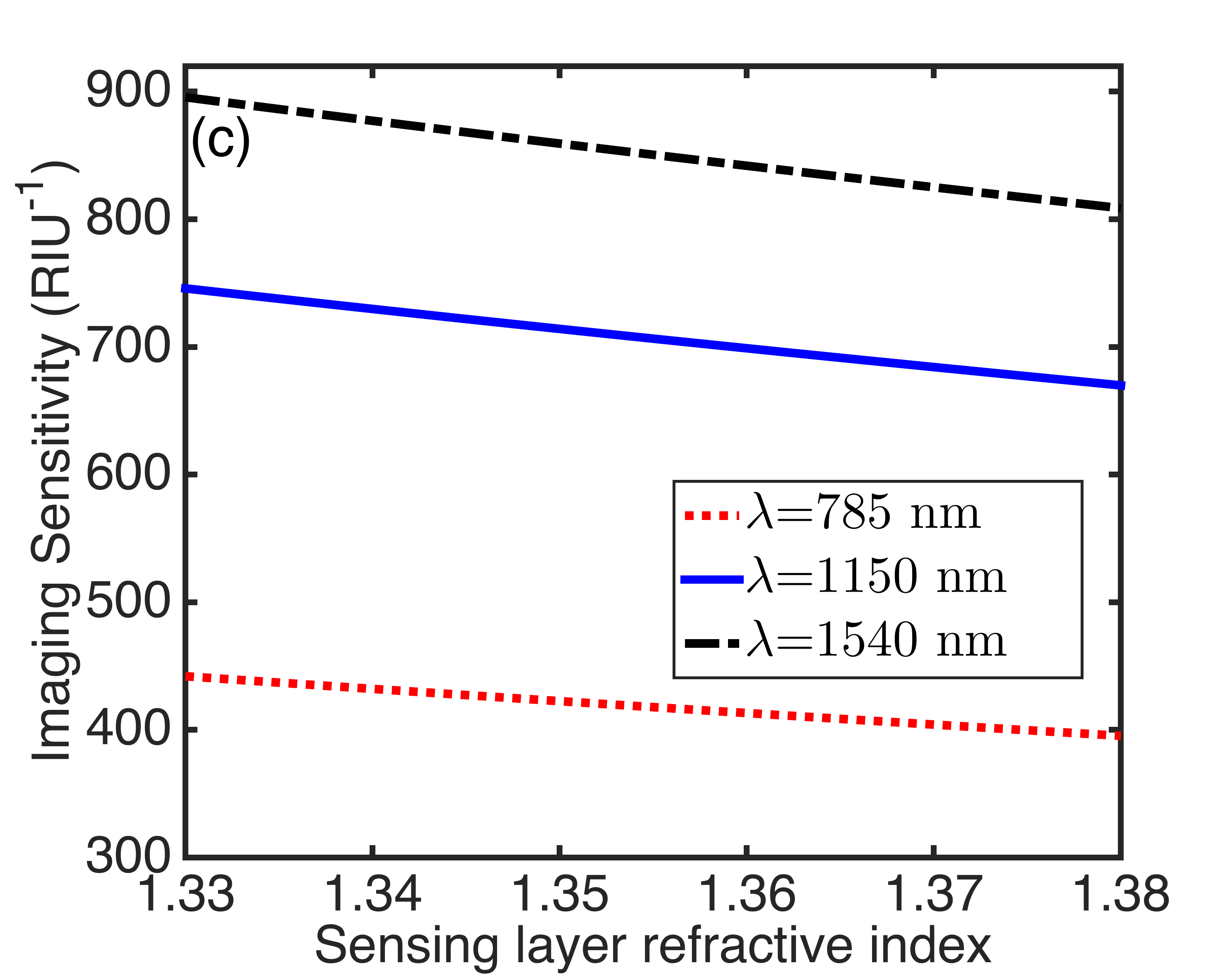}\\       
      \caption{(a) Resonance angle, (b) FWHM, and (c) imaging sensitivity as a function of the sensing layer refractive index at three different wavelengths, $\lambda=785$ nm, 1150nm and 1540 for monolayer $\text{MoS}_2$ based imaging sensor. The Al thickness is 38 nm.}
\label{fig6}
\end{figure}

The sensing layer RI is another important parameter for the sensor performance. As shown in Fig. \ref{fig6}(a), the resonance angle increases with the sensing layer RI, which can be understand from the SPR resonance condition \cite{homola2008surface}:
\begin{equation}
\frac{2 \pi}{\lambda} n_p \sin\theta_{\text{SPR}} = \text{Re}\left\{ \frac{2 \pi}{\lambda} \left({\frac{\epsilon_{m}n_s^2}{\epsilon_{m}+n_s^2}}\right)^{1/2} \right\},
\label{SPR}
\end{equation} 
where $n_p$ is the RI of prism, $\epsilon_{m}$ is the dielectric constant of metal film. The left-hand side of Eq. (\ref{SPR}) is the propagation constant of the incident light, and the term on the right hand side is the  real part of propagation constant of SPW. The propagation constant of the SPW increases with the RI of sensing layer, which leads to the increase in the resonance angle. In addition, the resonance angle also increases with the incident wavelength (Fig. \ref{fig6}(a)). For the sensor performance, the FWHM increases and the imaging sensitivity decreases with the sensing layer RI (see Fig. \ref{fig6}(b) and (c)).

\begin{figure}[thpb]
      \centering
      \includegraphics[scale=0.18]{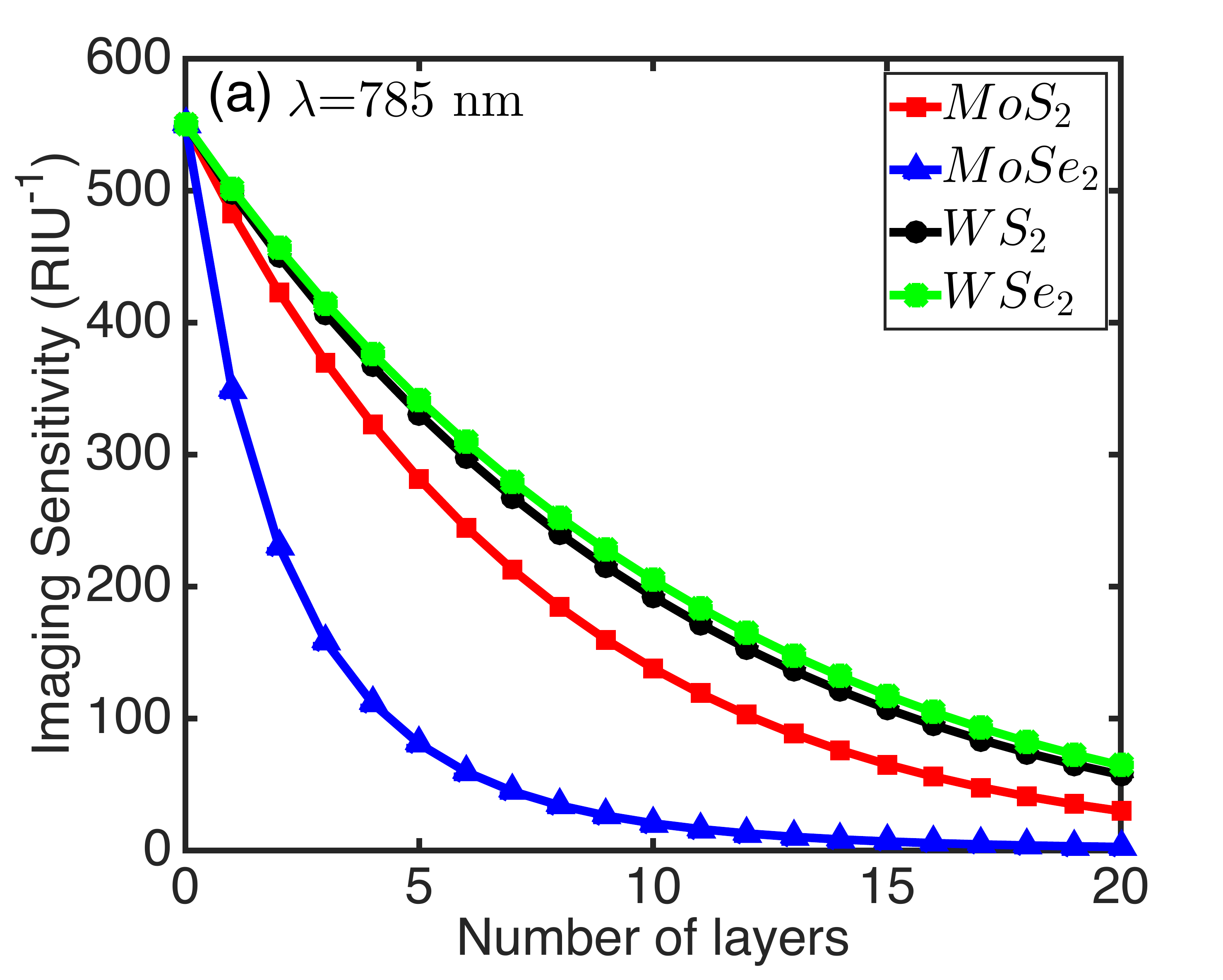}\\
      \includegraphics[scale=0.18]{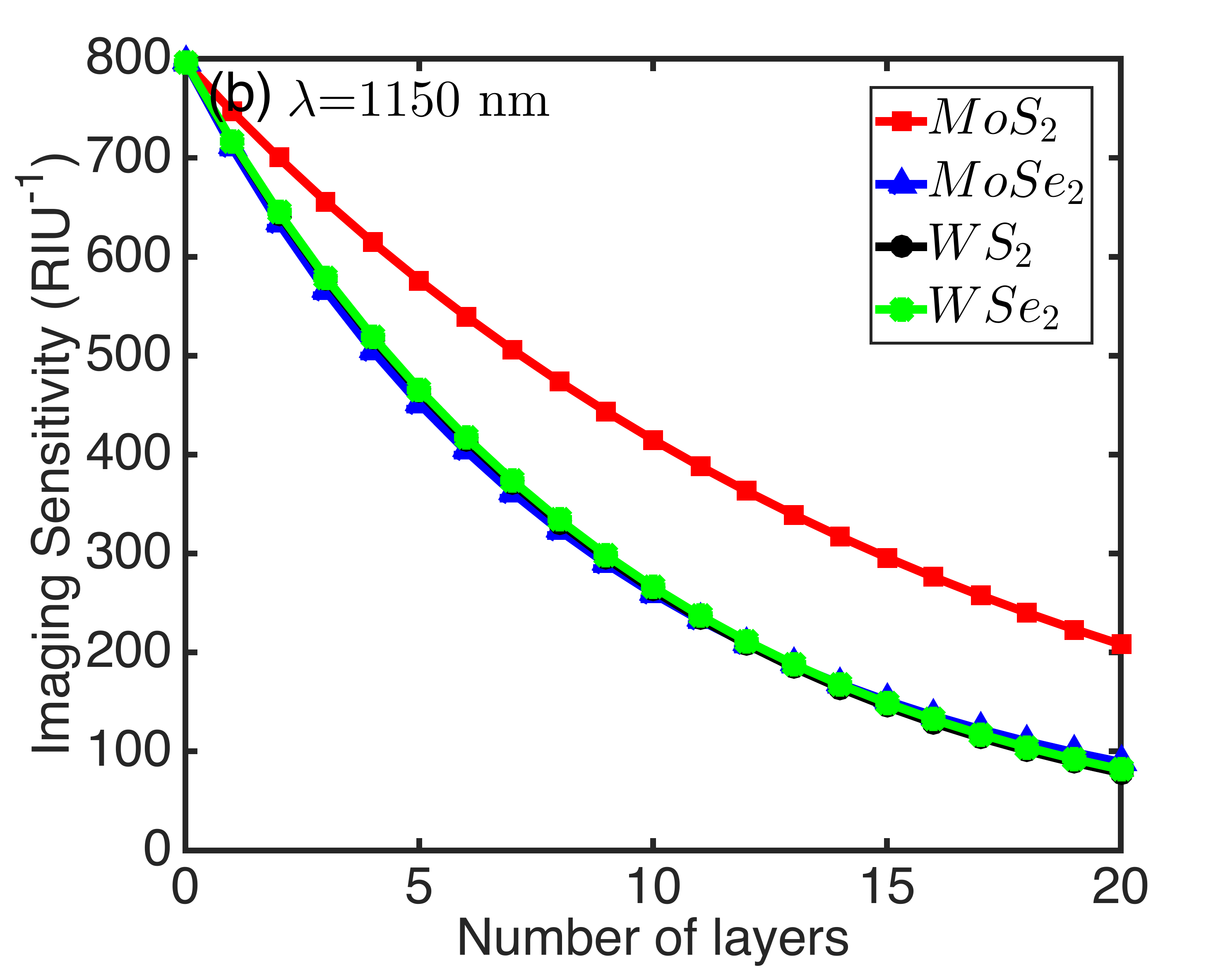}\\
      \includegraphics[scale=0.18]{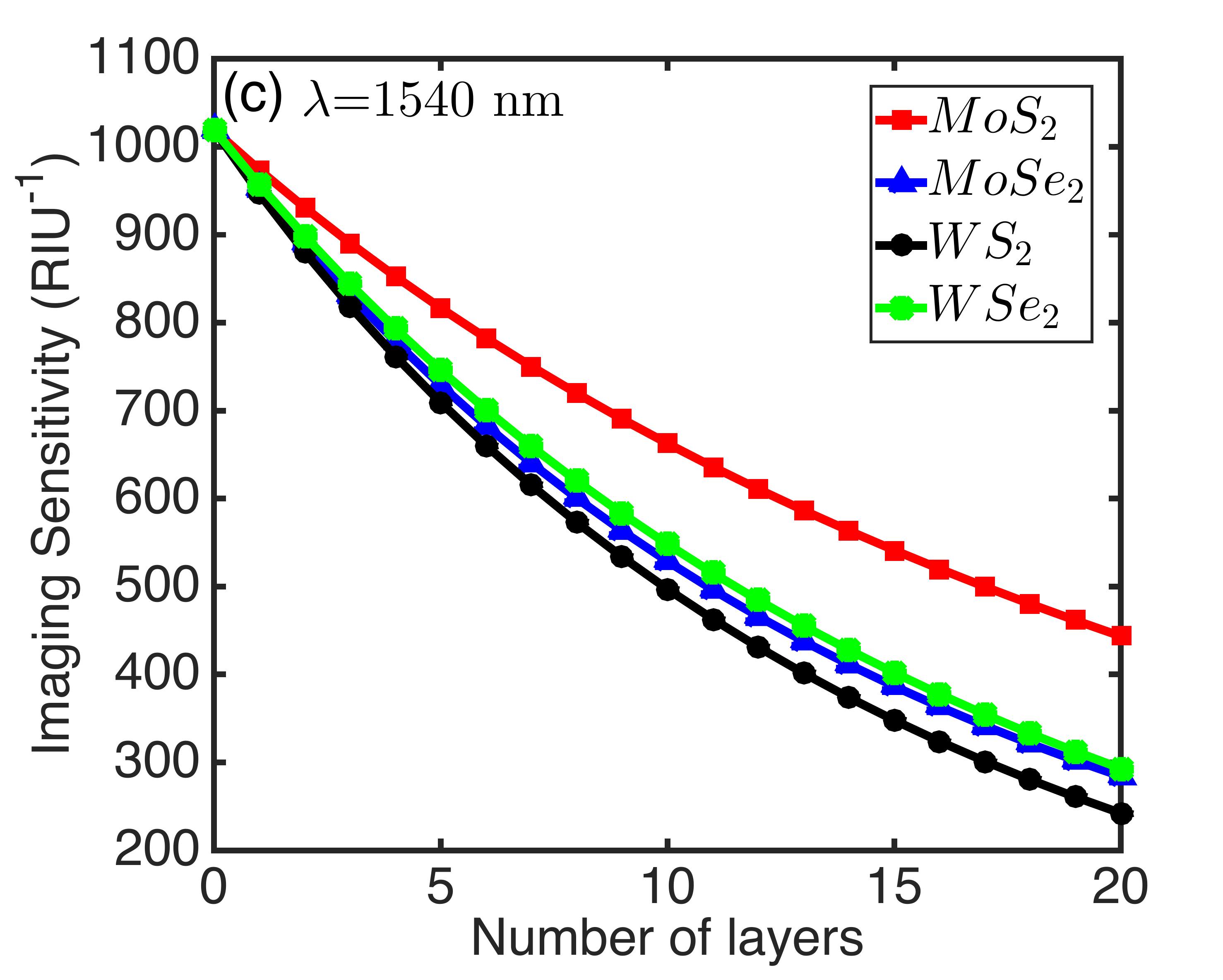} \\     
      \caption{Imaging sensitivity for SPR imaging sensor with multiple TMDC layers. (a) $\lambda=785$ nm with 42.8 nm thick Al film, (b) $\lambda=1150$ nm with 37.6 nm thick Al film, and (c) $\lambda=1540$ nm with 33.4 nm thick Al film.}
\label{fig7}
\end{figure}

\begin{figure}[thpb]
      \centering  
      \includegraphics[scale=0.18]{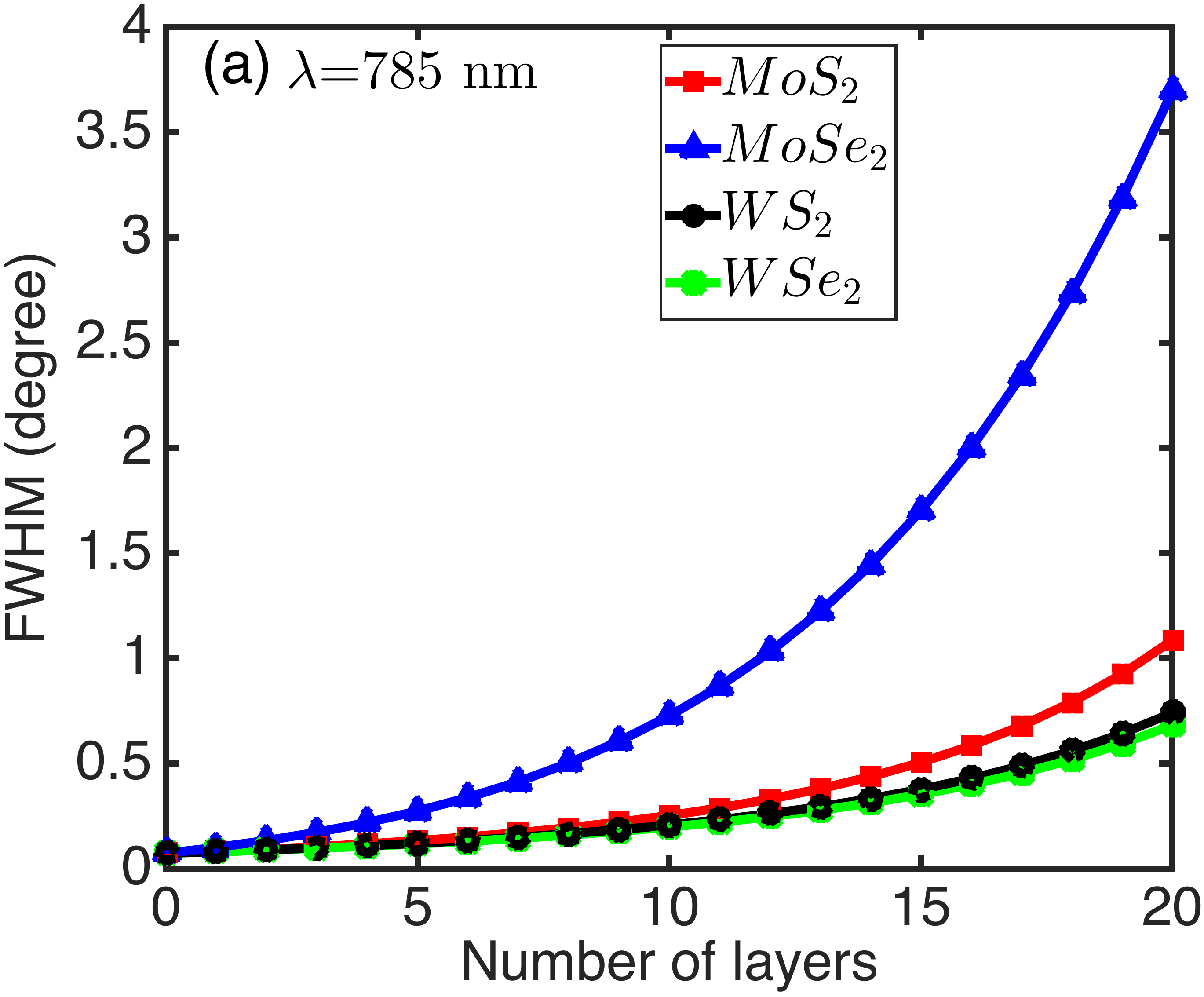}\\
      \includegraphics[scale=0.18]{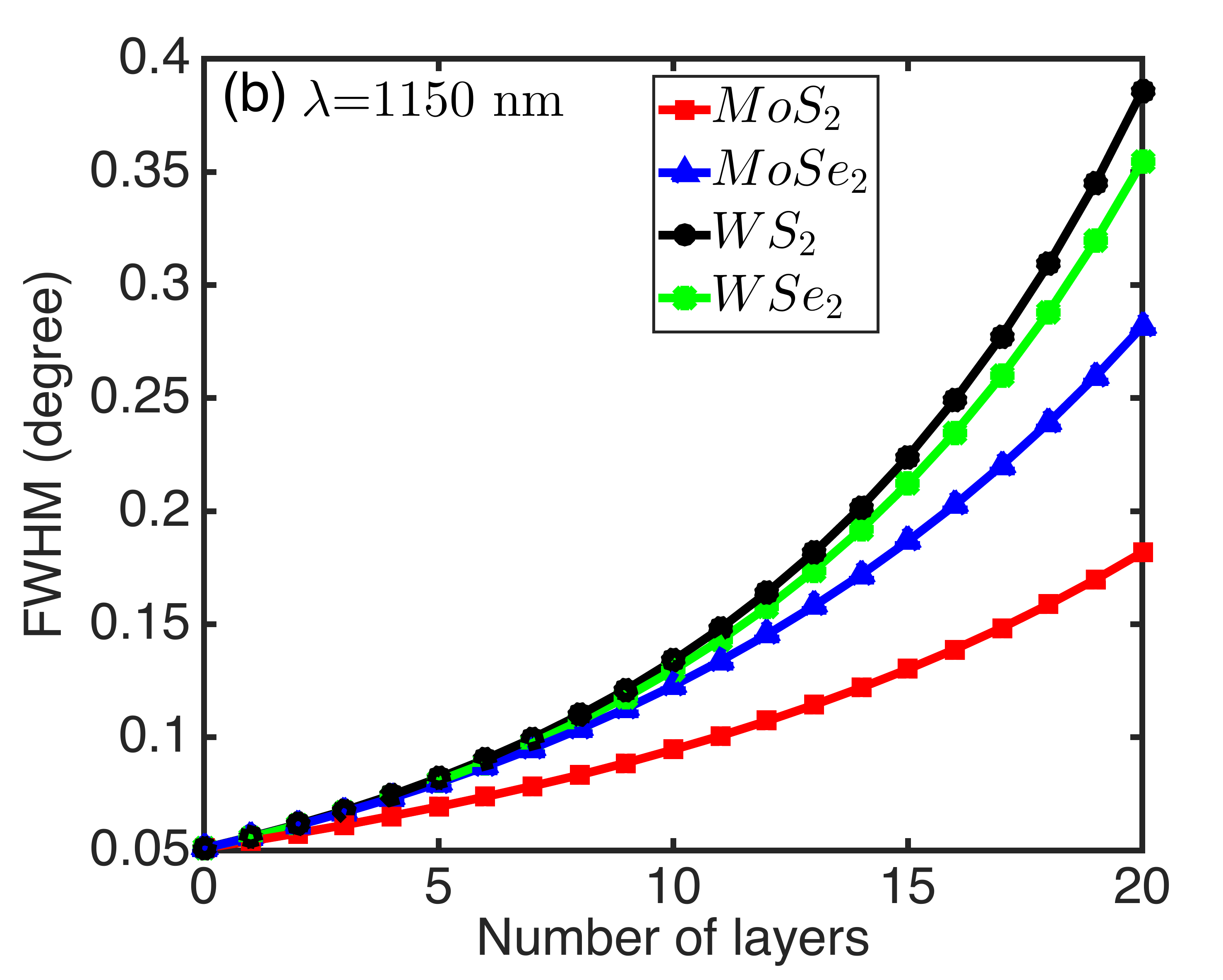}\\
      \includegraphics[scale=0.18]{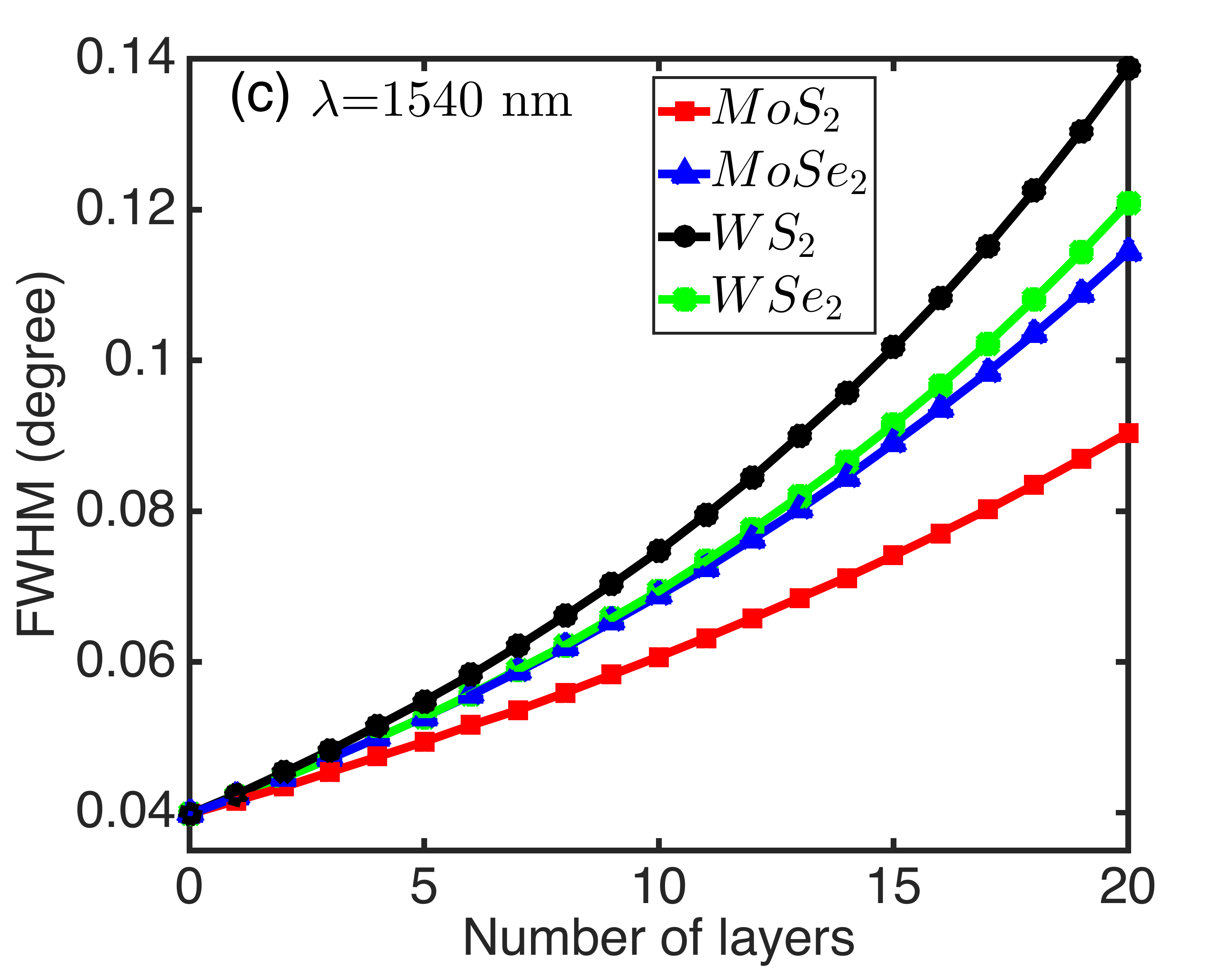}\\         
      \caption{FWHM for SPR imaging sensor with multiple TMDC layers. (a) $\lambda=785$ nm with 42.8 nm thick Al film, (b) $\lambda=1150$ nm with 37.6 nm thick Al film, and (c) $\lambda=1540$ nm with 33.4 nm thick Al film.}
\label{fig8}
\end{figure}

Other 2D TMDC materials, like Molybdenum diselenide ($\text{MoSe}_2$), Tungsten disulfide ($\text{WS}_2$) and Tungsten diselenide ($\text{WSe}_2$), have been proposed and demonstrated for sensing applications \cite{ouyang2017two,ouyang2016sensitivity,yang2017gas,yang2015graphene}. It is interesting to compare the sensor performance of SPR imaging sensor with different TMDC materials. Here, we compare the imaging sensitivity and FWHM for SPR imaging sensor with four TMDC materials ($\text{MoS}_2$, $\text{MoSe}_2$, $\text{WS}_2$, and $\text{WSe}_2$) at three different wavelengths, as shown in Fig. \ref{fig7} and \ref{fig8}, respectively. The thickness of Al film are taken from Fig. \ref{fig3} that have been optimized for the three wavelengths: 42.8 nm for $\lambda=785$ nm, 37.6 nm for $\lambda=1150$ nm and 33.4 nm for $\lambda=1540$ nm. Similar to the $\text{MoS}_2$-based SPR imaging sensor, imaging sensitivity decreases while FWHM increases with the number of layers for $\text{MoSe}_2$-, $\text{WS}_2$-, and $\text{WSe}_2$-based imaging sensor. For $\lambda=785$ nm, $\text{WSe}_2$ exhibits the highest imaging sensitivity with minimum FWHM, whereas $\text{MoSe}_2$-based sensor has the smallest sensitivity and maximum FWHM. At $\lambda=1150$ nm, $\text{MoS}_2$ SPR imaging sensor has the maximum sensitivity as well as the minimum FWHM, same as the case of $\lambda=1540$ nm. For the other three TMDC materials at wavelengths $\lambda=1150$ nm and $\lambda=1540$ nm, they exhibit similar imaging sensitivity, and $\text{WS}_2$-based SPR imaging sensor has the maximum FWHM. Therefore, to obtain a high performance SPR imaging sensor, one can choose $\text{WSe}_2$-based sensor at $\lambda=785$ nm, while for wavelengths $\lambda=1150$ nm and $\lambda=1540$ nm, $\text{MoS}_2$ is a better choice. 

\section{Conclusion}
In this work, an ultrahigh sensitive SPR imaging biosensor based on $\text{MoS}_2$-on-Al is proposed. In the designed biosensor structure, $\text{MoS}_2$ layers are employed to inhibit the oxidation of Al thin film and as the recognition layer to capture biomolecules. Graphene-on-Al based sensor exhibits better sensor performance than that of the proposed sensor in the visible range, however, the $\text{MoS}_2$-based SPR imaging biosensor overtakes its graphene counterparts in the near-infrared regime. It is found that the imaging sensitivity decreases with the number of $\text{MoS}_2$ layers applied, while the FWHM increases. A similar trend is observed with the RI of sensing layer. In addition, better sensor performance can be obtained at higher wavelength, where the imaging sensitivity can be as high as $\sim$974 $\text{RIU}^{-1}$ at wavelength $\lambda=1540$ nm. Compared with other TMDC materials ($\text{MoSe}_2$, $\text{WS}_2$, and $\text{WSe}_2$) based biosensor, biosensor based on $\text{WSe}_2$ shows the best sensor performance at wavelength $\lambda=785$ nm, while $\text{MoS}_2$-based biosensor has better performance than the other three biosensors at wavelengths $\lambda=1150$ nm and $\lambda=1540$ nm. We believe that the present study will be helpful in designing a high performance SPR imaging sensor for chemical and and biosensing applications.  

\section*{Funding}
Singapore ASTAR AME IRG (A1783c0011). 

\end{document}